\documentclass[a4paper]{article}

\usepackage{Odyssey2020}
\usepackage{epsfig,amssymb,amsmath}
\usepackage{multirow}
\usepackage{hyperref}
\ninept

\newcommand{\vect}[1]{\boldsymbol{\mathrm{#1}}}
\newcommand{\matr}[1]{\boldsymbol{\mathrm{#1}}}
\newcommand{\norm}[1]{\left\lVert#1\right\rVert}

\title{Deep Speaker Embeddings for Far-Field Speaker~Recognition on Short Utterances}

\name{\begin{tabular}{c} Aleksei Gusev$^{1,2}$, Vladimir Volokhov$^2$, Tseren Andzhukaev$^2$, Sergey Novoselov$^{1,2}$, Galina Lavrentyeva$^{1,2}$, \\ Marina Volkova$^{1,2}$, Alice Gazizullina$^{1,2}$, Andrey Shulipa$^1$, Artem Gorlanov$^2$, Anastasia Avdeeva$^2$, \\ Artem Ivanov$^2$, Alexander Kozlov$^2$, Timur Pekhovsky$^2$, Yuri Matveev$^{1, 2}$\end{tabular}}

\address{  
\begin{tabular}{c}
  $^1$ITMO University, St. Petersburg, Russia\\
  $^2$STC-innovations Ltd., St. Petersburg, Russia\\
  \\
\small \tt {\{gusev-a,volokhov,andzhukaev,novoselov,lavrentyeva,volkova,gazizullina,} \\ 
\small \tt {shulipa,gorlanov,avdeeva-a,ivanov-ar,kozlov-a,tim,matveev\}@speechpro.com } 
\end{tabular}
}

\begin{document}

\maketitle

\begin{abstract}
Speaker recognition systems based on deep speaker embeddings have achieved significant performance in controlled conditions according to the results obtained for early NIST SRE (Speaker Recognition Evaluation) datasets. From the practical point of view, taking into account the increased interest in virtual assistants (such as Amazon Alexa, Google Home, Apple Siri, etc.), speaker verification on short utterances in uncontrolled noisy environment conditions is one of the most challenging and highly demanded tasks. This paper presents approaches aimed to achieve two goals: a) improve the quality of far-field speaker verification systems in the presence of environmental noise, reverberation and b) reduce the system quality degradation for short utterances. 
For these purposes, we considered deep neural network architectures based on TDNN (Time Delay Neural Network) and ResNet (Residual Neural Network) blocks. We experimented with state-of-the-art embedding extractors and their training procedures. Obtained results confirm that ResNet architectures outperform the standard x-vector approach in terms of speaker verification quality for both long-duration and short-duration utterances. We also investigate the impact of speech activity detector, different scoring models, adaptation and score normalization techniques. 
The experimental results are presented 
for publicly available data and verification protocols for the VoxCeleb1, VoxCeleb2, and VOiCES datasets.

\end{abstract}

\section{Introduction}
The increasing interest in reliable means of guarding and restricting access to informational resources requires the development of new authentication methods. Biometric recognition remains one of the priority research areas in this field.

Today Automatic Speaker Verification (ASV) systems are a subject of increased interest of both state law enforcement agencies and commercial structures due to their reliability, convenience, low cost and provided security. Moreover, such systems can operate on different input-output devices and communication channels (landline, mobile
telephone networks, IP telephony, etc.).

The latest results obtained for the telephone part of NIST SRE (National Institute of Standards and Technology Speaker Recognition Evaluation) datasets demonstrated that Speaker Recognition (SR) systems based on deep speaker embeddings had achieved significant results in controlled conditions \cite{VCSGRMSBGPRDTCD2020}. However, speaker verification on short utterances is still one of the more challenging tasks in the text-independent speaker recognition field.

Taking into account the increased interest in virtual assistants (such as Amazon Alexa, Google Home, Apple Siri, etc.), the demand for far-field speaker verification on short utterances (such as wake-up words and short commands) in uncontrolled noisy environment conditions is very high. 

Such factors as channel mismatch, environmental noise and room reverberation can dramatically decrease the quality of these systems. This was confirmed by the VOiCES from a Distance challenge 2019 (VOiCES~2019 challenge) \cite{RBABFGLNSHGHSN2018, NHMRLB2019} aimed to support research in the area of speaker recognition and automatic speech recognition with the special focus on single channel far-field audio under noisy conditions.

This paper presents approaches aimed to achieve two goals simultaneously: to improve the performance of far-field speaker verification systems in the presence of environmental noise and reverberation, and to reduce the system quality degradation for short utterances.
In order to achieve these goals, we consider state-of-the-art deep neural network architectures and its applicability for speaker verification task in uncontrolled environmental conditions on publicly available data and verification protocols for the VoxCeleb1, VoxCeleb2, and VOiCES datasets.

We experimented with deep speaker embedding extractors based on TDNN (Time Delay Neural Network) \cite{SGRPK2017} and ResNet (Residual Neural Network) \cite{VCSGRMSBGPRDTCD2020, ZWSMP2019} blocks and different training objectives. A detailed description of the extractors is presented in Section~\ref{sec:implementation}.
Special attention was paid to the impact of deep neural network speech activity detector presented in \ref{sec:vad} that is more robust against noise and other distortions compared to classical energy-based methods.
In this paper, we also analyzed different scoring models, adaptation and score normalization techniques and estimated their contribution to the final system performance.

All obtained experimental results and their comparison with the standard x-vector approach are considered in Section \ref{sec:res}. The proposed systems performance is presented in terms of EER (Equal Error Rate) and minDCF (Minimum Detection 
Cost Function).

\section{Related work}
Implementation of deep learning approaches for speaker representation undoubtedly lets the speaker recognition field reach new levels of its evolution. Latest trends in the deep learning area applied to the speaker recognition problem form new state-of-the-art SR systems.

\subsection{DNN speaker embeddings}
Deep neural network based speaker embedding extractors substantially improve the performance of speaker ID systems in challenging conditions.
TDNN based x-vector system significantly outperformed conventional i-vector based system in terms of speaker recognition performance and hence became new baseline for text-independent SR task \cite{SGRPK2017}.
The authors proposed an end-to-end system that learns to classify speakers and produce representative deep speaker embeddings able to generalize well to speakers that have not been seen in the training data. The key feature of the proposed architecture was a statistics pooling layer designed to accumulate speaker information from the whole speech segment into one -- x-vector. Extracted from an intermediate layer of the neural network which comes after the statistics pooling layer, x-vectors demonstrate properties similar to those of i-vectors from total variability space, which makes it possible to effectively use them in the standard Linear Discriminant Analysis (LDA) followed by Probabilistic Linear Discriminant Analysis (PLDA) \cite{K2010} backend.

Studies such as \cite{SGRSMPK2019, NSKKS2018} follow this deep speaker representation direction with improvement of SR performance. For example, the system from \cite{SGRSMPK2019} proposed by JHU team for NIST SRE 2018 used the extended version of TDNN based architecture -- E-TDNN. The differences include an additional TDNN layer with wider temporal context and unit context TDNN layers between wide context TDNN layers. 

Paper \cite{NSKKS2018} proposes to use an alternative training objective -- A-Softmax (Angular Margin Softmax) activation \cite{LWYLRS2017} -- instead of the standard Softmax to train a so called c-vector based system.
The main characteristics of the proposed architecture were residual blocks  \cite{HZRS2016}  built using  TDNN  architecture and MFM (Max-Feature-Map) activations \cite{LNMKKS2017} used instead of ReLU.

\subsection{Speaker embeddings for short utterances}
Short utterances and far-field microphones are new challenging conditions for the SR task. Recent papers \cite{XNCZ2019, HE2019} devoted to this problem demonstrate that substantial improvements can be achieved by deeper architectures such as residual networks \cite{HZRS2016} and by more accurate task-oriented augmentation of training data.

An analysis of the degradation of speaker verification quality at short intervals on the VoxCeleb1 dataset was carried out in \cite{XNCZ2019, HE2019}.
Authors of \cite{XNCZ2019} demonstrated impressive results for "in the wild" scenario. They proposed a modified residual network with a NetVLAD/GhostVLAD layer for feature aggregation along the temporal axis. This layer is aimed to apply self-attentive mechanism with learnable dictionary encoding \cite{CCZWL2018}. 

An alternative approach for feature aggregation over time in a residual network is discussed in \cite{HE2019}. The authors proposed a simple and elegant Time-Distributed Voting (TDV) method. It demonstrates significant quality improvement for short utterances in comparison with NetVLAD solution. However, it does not perform so well on longer duration utterances.

\subsection{Speaker embeddings for distant speaker~recognition}
Recent progress and growing popularity of virtual assistants in smart home systems and smart devices have led to higher requirements not only for speech recognition but for the reliability of the biometric systems under far-field conditions as well.
In 2019 the VOiCES from a Distance Challenge \cite{NHMRLB2019} was organised to support the research in the area of speaker recognition and automatic
speech recognition with the special focus on single channel distant/far-field audio under noisy conditions. The challenge was based on the freely-available Voices Obscured in Complex Environmental Settings (VOiCES) corpus \cite{RBABFGLNSHGHSN2018} released several months before. 
Almost all systems proposed during the challenge exploited different architectures of neural networks to obtain deep speaker representations.
To reduce the effects of room reverberation and various kinds of distortions, some researches use more accurate task-oriented data augmentation \cite{NGIPSLVK2019, DKU_voices, BUT_voices, JHU_voices} and speech enhancement methods \cite{DKU_voices} based on single-channel weighted prediction error (WPE) \cite{WPE}.

\subsection{Loss function for speaker embedding learning}
Over the past few years, in the face recognition field, many loss functions have been proposed for the effective training of embedding extractors: A-Softmax \cite{LWYLRS2017}, AM-Softmax (Additive Margin Softmax) \cite{WLC2018}, AAM-Softmax (Additive Angular Margin Softmax) \cite{DGXZ2019}, D-Softmax (Dissected Softmax) \cite{HWLW2019} based loss functions. 
Recent studies in speaker verification field demonstrated impressive performance of the AM-Softmax based training loss function for speaker ID systems \cite{VCSGRMSBGPRDTCD2020, ZWSMP2019}.
Thus in this work, we mainly focused on the well-performing AM-Softmax based loss function and additionally experimented with D-softmax loss.

AM-Softmax based loss function is defined as follows:

\begin{equation}
\begin{split}
\mathcal{L}=-\frac{1}{N}\sum_{i}\frac{e^{s\left(cos\left(\theta_{y_i}\right)-m\right)}}{e^{s\left(cos\left(\theta_{y_i}\right)-m\right)}+\sum_{j\neq y_i}e^{s\left(cos\left(\theta_{j}\right)\right)}},
\label{ams_loss}
\end{split}
\end{equation}
where $\cos\left(\theta_{y_i}\right)=\mathbf{w}_{y_i}^{T}\mathbf{f}_i/\left(\left\|\mathbf{w}_{y_i}\right\|\|\mathbf{f}_{i}\right\|)$, $\mathbf{w}_{y_i}$ is the weight vector of class $y_i$, and $\mathbf{f}_i$ is the input to the layer $i$. Parameter $s$ is an adjustable scale factor and $m$ is the penalty margin. AM-Softmax loss allows to compare speaker embeddings by cosine distance.

D-Softmax based loss is a new loss function that was presented recently in \cite{HWLW2019} as an effective objective for face embedding learning. Authors of \cite{HWLW2019} speculate that the intra- and inter-class objectives in the categorical cross entropy loss are entangled, therefore a well-optimized inter-class objective leads to relaxation on the intra-class objective, and vice versa. The main idea of D-Softmax based loss is to dissect the cross entropy loss into independent intra- and inter-class objective.

D-Softmax based loss function is defined as follows:

\begin{equation}
\begin{split}
&\mathcal{L}=\mathcal{L}_{intra}+\mathcal{L}_{inter}= \\
&-\frac{1}{N}\sum_{i}\left(\frac{e^{s\cos\left(\theta_{y_i}\right)}}{e^{s\cos\left(\theta_{y_i}\right)}+\epsilon}+\frac{1}{{1}+\sum_{j\neq y_i}e^{s\cos\left(\theta_{j}\right)}}\right),
\label{ds_loss}
\end{split}
\end{equation}
where $\epsilon$ and $s$ are customizable parameters.

\section{Description of the system components}
\subsection{Feature extraction}

For all our embedding extractors we used MFCC (Mel Frequency Cepstral Coefficients) and MFB (Log Mel-filter Bank Energies) from 16 kHz raw input signals (standard Kaldi recipe) as low-level features:
\begin{itemize}
\item 40 dimensional MFCC extracted from the raw signal with 25ms frame-length and 15ms overlap;
\item 80 dimensional MFB extracted from the raw signal with 25ms frame-length and 15ms overlap.
\end{itemize}

For extracted voice features we applied 2 different postprocessing techniques depending on the type of embedding extractor used afterwards:
\begin{itemize}
    \item local CMN-normalization (Cepstral Mean Normalization) over a 3-second sliding window;
    \item local CMN-normalization over a 3-second sliding window and global CMVN-normalization (Cepstral Mean and Variance Normalization) over the whole utterance.
\end{itemize}

For our neural network based VAD solution we used MFCC features extracted from signal downsampled to 8 kHz. The detailed description is presented below. 

\subsection{Voice activity detection}
\label{sec:vad}
Besides energy-based VAD (Voice Activity Detector) from Kaldi Toolkit and ASR based VAD \cite{MKRSMBAPKPZ2019} in this work we investigated our new neural network based VAD. 

This work adapts the U-net \cite{RFB2015} architecture to the task
of speech activity detection. Such architecture was originally introduced in biomedical imaging for semantic segmentation in order to improve precision and localization of microscopic images. It builds upon the fully convolutional network 
and is similar to the deconvolutional network 
In a deconvolutional network, a stack of convolutional layers -- where each layer halves the size of the image but doubles the number of channels -- encodes the image into a small and deep representation. That encoding is then decoded to the original size of the image by a stack of upsampling layers. 

\begin{figure}[h]
  \centering
  \includegraphics[width=1.0\linewidth]{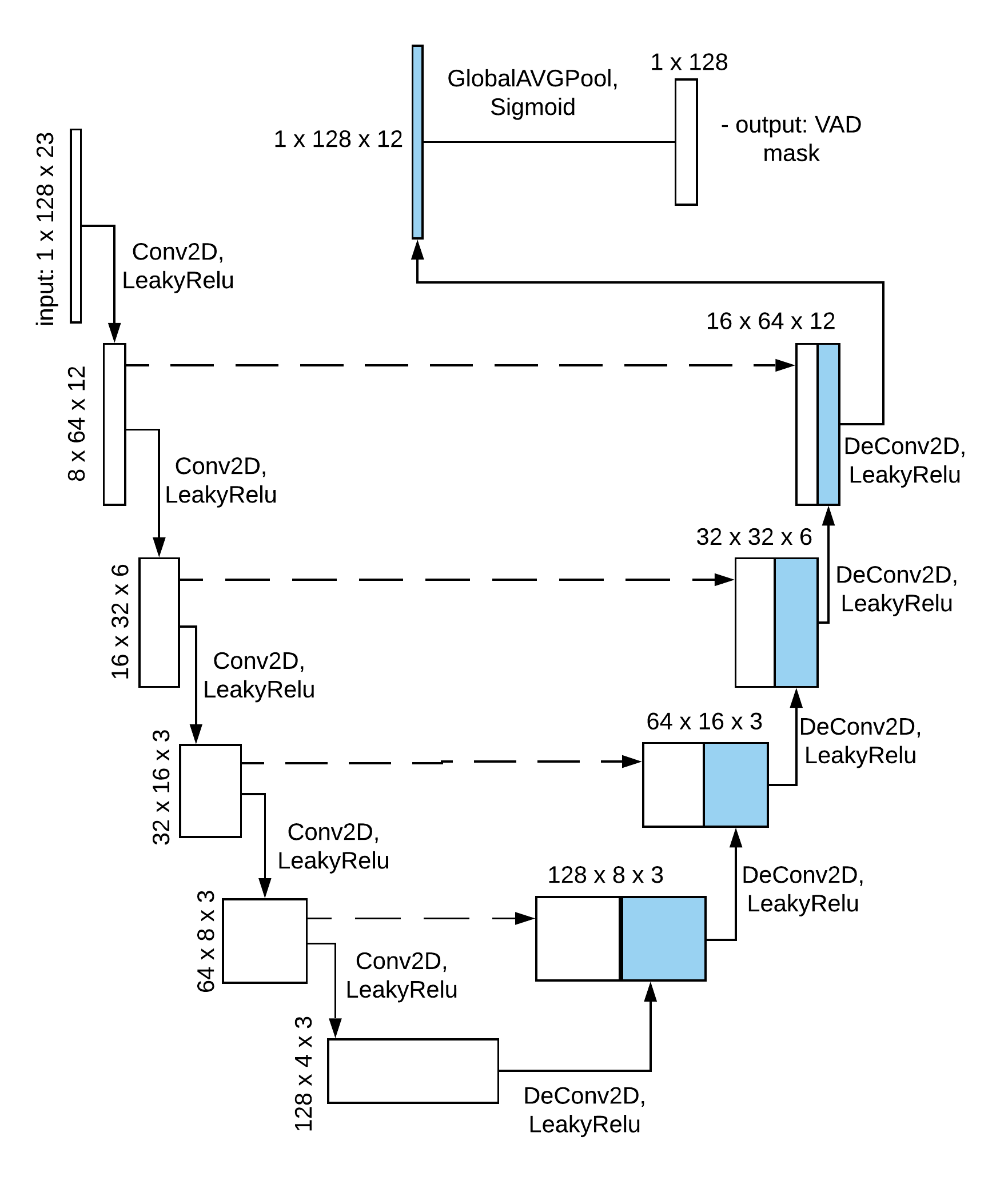}
  \caption{{\it U-net based VAD architecture}}
  \label{fig:unetvad}
\end{figure}

Our U-net based VAD is built on a modified and reduced version of the original architecture. Figure~\ref{fig:unetvad} schematically outlines the proposed version of neural network. 
It takes 8kHz 23-dimensional MFCC as input features. Our VAD solution works with a half overlapping 2.56 sec sliding window and a 1.28sec overlap. It should be noted that each MFCC vector is extracted for 25ms frame every 20ms. 
This results in  $128\times23$ input features size for the neural network. 

The goal of the neural network is to predict the 128 dimensional speech activity mask for every 2.56sec speech segment. Thus the resolution of the proposed speech detector is equal to 20ms. 
The final decoder layer is a sigmoid activated global average pooling layer. Its output is used as the speech activity mask.

The U-net is trained on artificially augmented data with speech labels obtained from the oracle handmade segmentation or using oracle ASR based VAD processing of clean version of the data. 

To train the network, we used a combination of binary cross entropy loss function and dice loss \cite{vnet_dice_loss}. The latter aims to maximize the dice coefficient between predicted binary segmentation set $p_i\in{P}$ and ground truth binary labels set $g_i\in{G}$:

\begin{equation}
\mathcal{D}=  \frac{2\sum_{i}^{N}p_ig_i}{\sum_{i}^{N}p_i^2 + \sum_{i}^{N}g_i^2},\\
\end{equation}
where the sums run over the $N$ frames.

\subsection{Embedding extractors}
We considered deep speaker embedding extractor with the most popular residual network architecture named ResNet34 and a deeper ResNet50 network \cite{HZRS2016}.

Table~\ref{table1} describes ResNet34 architecture we used. The key block of ResNet34 is ResNetBlock. It consists of two convolutional layers with $3\times3$ filters. ReLU activation follows each convolutional layer, and Maxout activation is used for embedding extraction. We apply batch normalization technique to stabilize and speed up network convergence. The settings for ResNet34 embedding extractors training were borrowed from \cite{ZWSMP2019}.

\begin{table}[th]
\caption{\label{table1} {\it Embedding extractor based on ResNet34 architecture configuration.}}
\vspace{2mm}
\centerline{
\begin{tabular}{|c|c|c|}
\hline
layer name & structure & output \\
\hline  \hline
Input         & 80 MFB log-energy & $80\times200\times{1}$ \\
Conv2D-1      & $3\times3$, stride 1 & $80\times200\times32$\\
ResNetBlock-1 &  $\begin{bmatrix} 3\times3, 32\\ 3\times3, 32 \end{bmatrix}\times3$, stride 1 & $80\times200\times32$\\
ResNetBlock-2 &  $\begin{bmatrix} 3\times3, 64\\ 3\times3, 64 \end{bmatrix}\times4$, stride 2 & $40\times100\times64$\\
ResNetBlock-3 &  $\begin{bmatrix} 3\times3, 128\\ 3\times3, 128 \end{bmatrix}\times6$, stride 2 & $20\times50\times128$\\
ResNetBlock-4 &  $\begin{bmatrix} 3\times3, 256\\ 3\times3, 256 \end{bmatrix}\times3$, stride 2 & $10\times25\times256$\\
StatsPooling  & mean and std & $20\times256$\\ 
Flatten       &  -- & 5120\\
Dense1        &  embedding layer & 512\\
Dense2        &  output layer & $\textit{N}_{spk}$\\
\hline
\end{tabular}}
\end{table}

More complex ResNet50 architecture contains three convolutional layers in ResNetBlock with $1\times1$, $3\times3$, and $1\times1$ masks. Additionally, we used SE (Squeeze-and-Excitation) blocks \cite{HSS2018} in each ResNetBlock.


\subsection{Backend}
In this work, we used Cosine Similarity (CS) and Cosine Similarity Metric Learning (CSML) for scoring. Additionally, adaptation and score normalization were applied.

\subsubsection{CS and CSML} We used CS to distinguish \textit{speaker embeddings}:

\begin{equation}
\mathcal{S(\vect{x_1},\vect{x_2})} = \dfrac{\vect{x_1}^T\vect{x_2}}{{\norm{\vect{x_1}}}{\norm{\vect{x_2}}}},
\end{equation}
where $(\vect{x_1}, \vect{x_2})$ are speaker embedding vectors.

As an alternative scoring model CSML approach 
was used for \textit{speaker verification}.
According to the original idea a linear transformation $\matr{A}$ was learned to compute cosine distance for a pair $(\vect{x_1}, \vect{x_2})$ as follows:

\begin{equation}
\label{csml}
\mathcal{S(\vect{x_1},\vect{x_2}, \matr{A})} = \dfrac{(\matr{A}\vect{x_1})^T(\matr{A}\vect{x_2})}{{\norm{\matr{A}\vect{x_1}}}{\norm{\matr{A}\vect{x_2}}}},
\end{equation}
where the transformation matrix $\matr{A}$ is upper triangular. However, unlike \cite{NB2010} the triplet loss objective function was used for $\matr{A}$ training.
%
The metric learning was performed similar to the way it was done in \cite{NSSKK2018} using TensorFlow framework.

\subsubsection{Domain adaption}
In this work, we used simple domain adaptation procedure \cite{ABK2018} based on centering on in-domain set (mean speaker embedding subtraction). The mean vector is calculated using adaptation set in this case.

\subsubsection{Score normalization}
Additionally, scoring systems normalization technique from \cite{CVDFKCL2017} was used.
For a pair $(\vect{x_1},\vect{x_2})$ the normalized score can be estimated as follows:

\begin{equation}
\label{eq:score_norm}
\mathcal{\hat{S}(\vect{x_1},\vect{x_2})} = \frac{\mathcal{S(\vect{x_1},\vect{x_2})}-\mu_1}{\sigma_1}+\frac{\mathcal{S(\vect{x_1},\vect{x_2})}-\mu_2}{\sigma_2},
\end{equation}
where the mean $\mu_1$ and standard deviation $\sigma_1$ are calculated by matching $\vect{x_1}$ against impostor cohort and similarly for $\mu_2$ and $\sigma_2$. A set of the $\textit{n}$ best scoring impostors were selected for each embedding pair when means and standard deviations are calculated.

\section{Implementation details}
\label{sec:implementation}
Here we describe speaker recognition systems and datasets used for their training.

\subsection{Datasets}
\label{sec:data}
In our experiment, we used three groups of training data:
\begin{itemize}
\item \textbf{TrainData-I} includes VoxCeleb1 \cite{NCZ2017} (without test data), VoxCeleb2 \cite{CNZ2018} and SITW \cite{MFCL2016} and their augmented versions. Augmentation was partially performed using standard Kaldi augmentation recipe (babble, music and noise) using the freely available MUSAN datasets\footnote{\label{openslr}\url{http://www.openslr.org}}.
Reverberation was performed using the impulse response generator based on \cite{AB1979}. Four different RIRs were generated for each of 40,000 rooms with a varying position of sources and destructors. It should be noted that, in contrast to the original Kaldi augmentation, we reverberated both speech and noise signals. In this case different RIRs generated for one room were used for speech and noise signals respectively. Thus we obtained more realistic data augmentation. We have already used this approach  
in our previous studies \cite{NGIPSLVK2019}. Energy-based VAD from Kaldi Toolkit was used to preprocess all samples from the database. The final database consists of approximately 5,200,000 samples (7,562 speakers);
\item \textbf{TrainData-II} contains VoxCeleb1Cat (without test data) and VoxCeleb2Cat (without test data) and their augmented versions. We concatenated all segments from the same session into one file.
Augmented data was generated using standard Kaldi augmentation recipe (reverberation, babble, music and noise) using the freely available MUSAN and RIR datasets\footnotemark{\ref{openslr}}. Energy-based VAD from Kaldi Toolkit was used to preprocess all samples from the database. The final database consists of approximately 830,000 samples (7,146 speakers);
\item \textbf{TrainData-III} is similar to TrainData-I, but ASR based VAD \cite{MKRSMBAPKPZ2019} was used to preprocess the examples from the database instead of the energy-based VAD;
\item \textbf{TrainData-IV} is similar to TrainData-II, but it contains only VoxCeleb2Cat (without test data) and its augmented version. The final database consists of approximately 727,800 samples (5,994 speakers).
\end{itemize}

\subsection{Extractors}
\textbf{ResNet34-MFB80-AM-TrainData-I:} This system is based on ResNet34 embedding extractor.
The key feature of this extractor is high dimensional input features (80 dimensional MFB).
Local CMN- and global CMVN-normalization are used to normalize extracted MFB features. This extractor was trained on short segments with the fixed 2 sec length and using AM-Softmax based loss.
Parameters $m$ and $s$ were respectively equal to 0.2 and 30 during the whole training stage. The learning rate was equal to 0.001 on the first two epochs, then it was decreased by a factor of 10 for each next epoch. TrainData-I was used for training. We trained this extractor for 4 epoch.

\textbf{Xvect-FTDNN-TrainData-I:} 
This system is based on the factorized TDNN embedding extractor \cite{VCSGRMSBGPRDTCD2020}. The main idea is that TDNN pre-pooling layers of the x-vector system are replaced by factorized TDNN with skip connections. Factorization of the weight matrix into two low-rank matrices, with one of them constrained to be semi-orthogonal, helps to reduce the number of neural network parameters. Using skip connections allows to solve the problem of gradient vanishing and makes training process more stable.
In our speaker embedding extractor, described in more detail in \cite{NGIPSLVK2019}, we slightly modified the original skip connections and reduced the size of TDNN layers.


\textbf{ResNet34-MFB80-D-TrainData-I:} This extractor is similar to ResNet34-MFB80-AM-TrainData-I, with the difference of using a fine-tuning procedure by means of D-Softmax loss function.

\textbf{ResNet50-SE-MFB80-AM-TrainData-I:} This system is based on ResNet50 embedding extractor with SE blocks. Input features, training procedure and etc. were equivalent to ResNet34-MFB80-AM-TrainData-I system.

\textbf{ResNet34-MFCC40-AM-TrainData-I:} This extractor is similar to ResNet34-MFB80-AM-TrainData-I, but uses 40 dimensional MFCC features as input. Local CMN- and global CMVN-normalization are applied.

\textbf{ResNet34-MFB80-AM-TrainData-II (2s):} This extractor is similar to ResNet34-MFB80-AM-TrainData-I, but was trained using TrainData-II dataset. AM-Softmax loss was used for the training with parameter $s$ equal to 30 during the whole training stage and parameter $m$ equal to 0.001 for the first epoch and to 0.2 for the next epochs. The initial value of learning rate was set to 0.001. The learning rate was decreased by a factor of 10 every next epoch. We trained this extractor for 4 epoch.

\textbf{ResNet34-MFB80-AM-TrainData-II (1s):}. This extractor is similar to ResNet34-MFB80-AM-TrainData-II, but it was trained using only 1 sec duration chunks. Each 100 MFB speech frames with no overlap extracted from all samples of the TrainData-II were used for training. AM-Softmax loss was used for training, parameter $s$ was equal to 30 during the whole training stage, parameter $m$ was set to 0.2 for all epochs. The learning rate was the same as in ResNet34-MFB80-AM-TrainData-II (2s) system.

\textbf{Xvect-Ext-TDNN-LSTM-TrainData-III:} This extractor is described in \cite{NGIPSLVK2019}. The system is the extended version \cite{SGRSMPK2019} of the original x-vector extractor, but with 9th layer replaced by LSTM-layer with cell dimension of 512, delay in the recurrent connections equal to -3, and both recurrent and non-recurrent projection dimension equal to 256. The LSTM layer context was reduced to 3. This embedding extractor was trained on TrainData-III.

\textbf{ResNet34-MFB80-AM-TrainData-IV (1s):} This extractor is similar to ResNet34-MFB80-AM-TrainData-II, but it was trained on 1 sec speech chunks obtained from TrainData-IV dataset in the same way as it was done for ResNet34-MFB80-AM-TrainData-II.

\section{Experiments and discussion}
\label{sec:res}

\subsection{Experimental setup}
All experiments described further in this paper were performed with the use of VoxCeleb1 \cite{NCXZ2019} and VOiCES~2019 challenge \cite{RBABFGLNSHGHSN2018, NHMRLB2019} datasets. The results are presented in terms of EER (Equal Error Rate) and minDCF (Minimum Detection Cost Function) for $P_{tar}=10^{-2}$ performance metrics.

\subsection{Preliminary investigation}
\label{subseg:prelim}

\begin{figure}[t]
\includegraphics[width=\columnwidth]{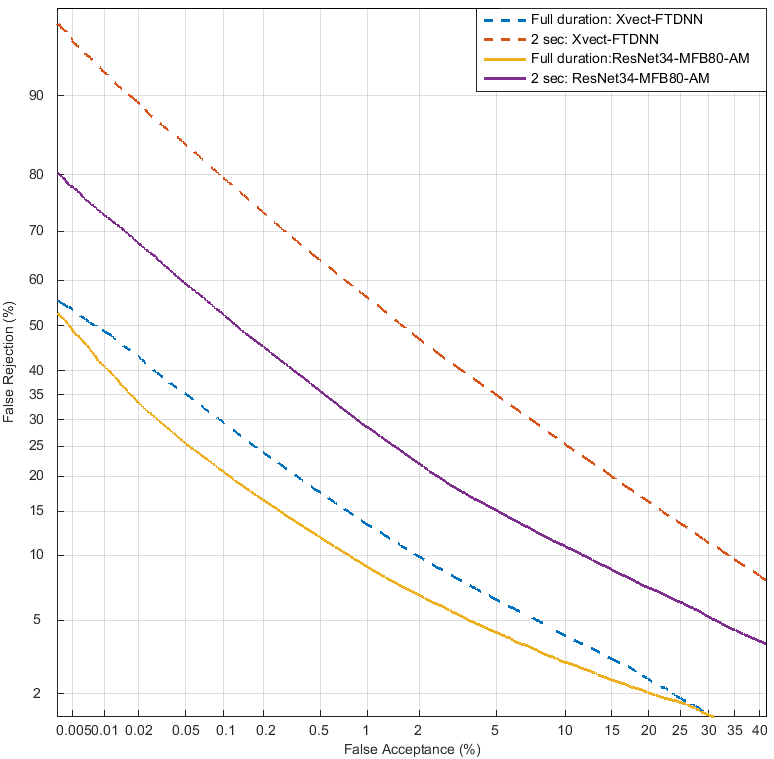}
\caption{{\it DET curves of Xvec-FTDNN-TrainData-I and ResNet34-MFB80-AM-TrainData-I embedding extractors for full and 2 sec duration of test utterances from VOiCES (eval set).}}
\label{fig:voices_eval}
\end{figure}


Our first goal was to investigate SR systems performance for the original full length testing protocols. Thus, Table \ref{tab:full_dur} demonstrates the experimental results obtained for the original "VOiCES dev", "VOiCES eval" and "VoxCeleb1-O cleaned" protocols.

We experimented with factorized TDNN based x-vector, ResNet34 and more complex ResNet50-SE networks. It should be noted that we used VOiCES eval part as a development (adaptation) set for the VOiCES dev part and vice versa. For the VoxCeleb1-O cleaned set we used a subset of 200k randomly selected clean files from VoxCeleb1 and VoxCeleb2 datasets in order to perform system adaptation and score normalization procedures. Top $200$ of the impostor scores were used to perform s-normalization. 
In these experiments we focused on different VAD and backend model configurations. We also experimented with the promising D-Softmax based loss function to improve system performance in case of ResNet34-MFB80-D-TrainData-I.

All results of the full duration experiments are presented in Table \ref{tab:full_dur}.



\begin{table*}[th]
\caption{\label{tab:full_dur} {\it Results of investigated systems for VOiCES (eval set), VOiCES (dev set) and VoxCeleb1-O (cleaned) protocols.}}
\vspace{2mm}
\centering
\resizebox{\textwidth}{!}{
\begin{tabular}{|c|c|cc|cc|cc|}
\hline
\multirow{2}{*}{Embedding extractor} & \multirow{2}{*}{Settings} & \multicolumn{2}{c|}{VOiCES (eval set)} & \multicolumn{2}{c|}{VOiCES (dev set)} & \multicolumn{2}{c|}{VoxCeleb1-O (cleaned)}                      \\ 
&    & minDCF & EER, \%  & minDCF & EER, \% & minDCF & EER, \%                                                        \\ \hline \hline
\multirow{4}{*}{ResNet34-MFB80-AM-TrainData-I} & CS backend                  & 0.366 & 5.31 & 0.155 & 1.30 & 0.193 & 1.96    \\ 
\multicolumn{1}{|l|}{}                      & + mean adapt.    & 0.327 & 4.88 & \textbf{0.144} & 1.32 & 0.195 & 1.89      \\
\multicolumn{1}{|l|}{}                      & + s-norm.    & 0.319 & 4.82 & 0.169 & 1.30 & 0.187 & 1.78    \\
\multicolumn{1}{|l|}{}                      & + U-net VAD          & \textbf{0.300} & \textbf{4.52} & 0.157 & \textbf{1.12} & 0.173 & 1.76   \\ \hline
\multirow{4}{*}{Xvect-FTDNN-TrainData-I}          & CSML backend                 & 0.496 & 7.12 & 0.253 & 2.28 & 0.363 & 4.15    \\
\multicolumn{1}{|l|}{}                      & + mean adapt.    & 0.426 & 6.03 & 0.234 & 1.99 & 0.381 & 4.18 \\
\multicolumn{1}{|l|}{}                      & + s-norm.    & 0.408 & 5.74 & 0.230 & 1.84 & 0.357 & 3.98 \\
\multicolumn{1}{|l|}{}                      & + U-net VAD          & 0.390 & 5.81 & 0.242 & 2.00 & 0.366 & 4.48  \\ \hline
ResNet34-MFB80-D-TrainData-I           & CS backend                  & 0.419 & 5.36 & 0.220 & 2.06 & 0.241 & 2.12 
\\
\hline
ResNet50-SE-MFB80-AM-TrainData-I                  & CS backend                   & 0.415 & 5.75 & 0.179 & 1.56 & 0.222 & 2.07    \\ 
\hline
ResNet34-MFCC40-AM-TrainData-I                     & CS backend                   & 0.405 & 6.03 & 0.178 & 1.38 & 0.236 & 2.14    \\ \hline
\multirow{4}{*}{ResNet34-MFB80-AM-TrainData-II}                & CS backend                   & 0.447 & 6.25 & 0.193 & 1.27 & 0.151 & 1.45    \\ 
\multicolumn{1}{|l|}{}                      & + mean adapt.    & 0.383 & 5.81 & 0.185 & 1.35 & 0.148 & 1.44      \\
\multicolumn{1}{|l|}{}                      & + s-norm.    & 0.366 & 5.78 & 0.201 & 1.42 & 0.152 & \textbf{1.39}    \\
\multicolumn{1}{|l|}{}                      & + U-net VAD          & 0.354 & 5.50 & 0.197 & 1.35 & \textbf{0.142} & 1.46 \\ \hline
Xvect-Ext-TDNN-LSTM-TrainData-III                      & \begin{tabular}[c]{@{}c@{}}CSML backend, \\ ASR VAD, \\ s-norm. \end{tabular} 
                                                                   & 0.349 & 5.16 & --    & --   & --    & --      \\ \hline
\end{tabular}
}
\end{table*}

\subsection{Results analysis of the full duration experiments}
Having analyzed the results obtained in \ref{subseg:prelim}, we can say that 
systems based on ResNet architectures outperform x-vector based systems in all our experiments.

One should note that U-net based VAD helps to improve the quality of systems for difficult conditions compared to the standard Kaldi energy-based VAD and S-normalization significantly improves the performance of all extractor types.

Our next observation is that appropriate training data preparation is an important step for system learning. 
Taking into account that TrainData-I (in contrast to TrainData-II) was prepared using special augmentation procedure (see Section \ref{sec:data}) to meet VOiCES dataset acoustic environment conditions it expectedly shows better performance for VOiCES test protocols. Thus comparison of ResNet34-MFB80-AM-TrainData-II and ResNet34-MFB80-AM-TrainData-I shows that despite the fact that both systems show good quality in various experiments, the more task-oriented training data preparation can significantly improve the quality of systems.

We should also note that increasing features resolution improves the quality of the systems. This can be approved by the results obtained for ResNet34-MFCC40-AM-TrainData-I and ResNet34-MFB80-AM-TrainData-I extractors.

From the results in Table \ref{tab:full_dur} one can see that the best performing system for VOiCES protocols is ResNet34-MFB80-AM-TrainData-I. It outperforms our previous best single system (Xvect-Ext-TDNN-LSTM-TrainData-I) submitted to the VOiCES challenge \cite{NGIPSLVK2019}.


The obtained results allow to conclude that D-Softmax based loss training does not help to improve ResNet34-MFB80 performance. We also did not achieve any improvement by using more complex ResNet50-SE based extractor in comparison with ResNet34. We suppose that it is caused by the more complex model being overfitted in this case.

For the VoxCeleb1-O (cleaned) protocol ResNet34-MFB80-AM-TrainData-II (2s) is the top performing system. It was trained on 2 sec speech chunks of TrainData-II dataset.

\subsection{Short utterance speaker recognition}
\label{sub:Main_experiments_p1}
In order to compare our SR systems performance for short utterances with those presented in \cite{XNCZ2019, HE2019} the special dataset was generated from the VoxCeleb1 corpus according to the description from \cite{XNCZ2019}: only files longer than 6 seconds (87010 utterances) were selected. A comparison protocol was generated by randomly sampling 100 target and 100 imposter pairs for each speaker from a total of 1,251 speakers in VoxCeleb1, resulting in 250,048 unique comparisons. We didn't succeed in obtaining the same size protocol as in \cite{XNCZ2019} because not all speakers had the necessary 100 target comparisons with samples longer than 6 seconds. But since the original model proposed in \cite{XNCZ2019} is freely available, we compared it with our ResNet34-MFB80-AM-TrainData-IV (1s) proposed above using the generated protocol. The comparison results of these models are shown in Table \ref{tab:short_dur_gen_proto}. The experiments were carried out in the same way as in \cite{XNCZ2019, HE2019} without using any VAD.
During our next experiments on short duration utterances, we used the following settings:
\begin{itemize}
    \item as enrollment samples, we used only full duration original files;
    \item as test samples, we used only the first 1, 2 and 5 seconds of speech in each file. If speech duration was less than required, we used all available speech in this file and didn't change the protocol. We applied CMN with a 3-second sliding window on these segments in the following way: 
    using VAD segmentation we accumulated the required amount of features, and applied CMN only to them discarding information redundant for the normalization;
    \item fixed protocol for all durations.
\end{itemize}

Table~\ref{tab:short_dur} demonstrates the comparison of all the described systems in case of different test samples duration (1~sec, 2~sec, 5~sec and full duration) for the same protocols in terms of EER.
In order to simplify the experiments and improve the reproducibility of our results, we did not use normalization and U-net based VAD for these tests.

Additionally, 
Fig.~\ref{fig:voices_eval} presents DET (Detection Error Tradeoff) curves of Xvec-FTDNN-TrainData-I and ResNet34-MFB80-AM-TrainData-I embedding extractors for full and 2~sec duration samples from 
VOiCES (eval set).


\begin{table*}[]
\caption{\label{tab:short_dur_gen_proto} {\it The results of the publicly available model from \cite{XNCZ2019} and the proposed ResNet34-MFB80-AM-TrainData-IV on the short utterances protocol (generated  from VoxCeleb1).}}
\vspace{2mm}
\centering
\begin{tabular}{|c|c|c|c|}
\hline
Embedding extractor & EER, \% (1s) & EER, \% (2s) & EER, \% (5s)                
\\ 
\hline \hline
Thin ResNet34 \cite{XNCZ2019}                  & 12.71 & 6.59 & 3.34
\\ 
\hline
ResNet34-MFB80-AM-TrainData-IV (1s)  & 9.91 & 4.48 & 2.26
\\ 
\hline
\end{tabular}
\end{table*}

\begin{table*}[h!]
\caption{\label{tab:short_dur} {\it Results of the investigated systems for VOiCES (eval set), VOiCES (dev set) and VoxCeleb1-O (cleaned) protocols in~relation to~different~length of test waveform.}}
\vspace{2mm}
\centering
\resizebox{\textwidth}{!}{
\begin{tabular}{|c|c|c|c|}
\hline
\multicolumn{1}{|c|}{Embedding extractor}                                                                 & \multicolumn{1}{c|}{\begin{tabular}[c]{@{}c@{}}VOiCES (eval set)\\ EER, \% (1s/2s/5s/full)\end{tabular}} & \multicolumn{1}{c|}{\begin{tabular}[c]{@{}c@{}}VOiCES (dev set)\\ EER, \% (1s/2s/5s/full)\end{tabular}} & \multicolumn{1}{c|}{\begin{tabular}[c]{@{}c@{}}VoxCeleb1-O (cleaned)\\ EER, \% (1s/2s/5s/full)\end{tabular}} \\ \hline \hline
\multicolumn{1}{|c|}{\begin{tabular}[c]{@{}c@{}}ResNet34-MFB80-AM-TrainData-II (2s), \\ CS backend\end{tabular}} & \multicolumn{1}{c|}{16.96/10.14/7.77/6.25}                                                               & \multicolumn{1}{c|}{8.70/\textbf{3.67}/\textbf{1.83}/\textbf{1.27}}                                                                & \multicolumn{1}{c|}{\textbf{6.77}/\textbf{2.74}/\textbf{1.59}/\textbf{1.45}}                                                                     \\ \hline
\multicolumn{1}{|c|}{\begin{tabular}[c]{@{}c@{}}ResNet34-MFB80-AM-TrainData-II (1s), \\ CS backend\end{tabular}} & \multicolumn{1}{c|}{19.85/14.27/12.61/11.17}                                                               & \multicolumn{1}{c|}{9.70/5.07/3.23/2.63}                                                                & \multicolumn{1}{c|}{6.87/3.18/2.11/1.99}                                                                     \\ \hline
\multicolumn{1}{|c|}{\begin{tabular}[c]{@{}c@{}}ResNet34-MFB80-AM-TrainData-IV (1s), \\  CS backend\end{tabular}}  & \multicolumn{1}{l|}{20.17/14.76/13.24/10.80} & \multicolumn{1}{c|}{10.13/5.15/3.27/2.46}  & \multicolumn{1}{c|}{7.13/3.54/2.23/2.09}   \\ \hline
\multicolumn{1}{|c|}{\begin{tabular}[c]{@{}c@{}}ResNet34-MFB80-AM-TrainData-I (2s), \\  CS backend\end{tabular}}   & \multicolumn{1}{c|}{16.26/\textbf{9.46}/\textbf{6.95}/\textbf{5.31}}    & \multicolumn{1}{c|}{8.97/4.15/1.97/1.30}   & \multicolumn{1}{c|}{8.04/3.47/2.08/1.96}   \\ \hline
\multicolumn{1}{|c|}{\begin{tabular}[c]{@{}c@{}}ResNet34-MFB80-AM-TrainData-I (1s), \\  CS backend\end{tabular}}  & \multicolumn{1}{c|}{\textbf{16.21}/10.63/9.20/8.00}   & \multicolumn{1}{c|}{\textbf{8.43}/4.35/2.57/2.06}   & \multicolumn{1}{c|}{7.51/4.03/2.60/2.57}   \\ \hline
\multicolumn{1}{|c|}{\begin{tabular}[c]{@{}c@{}}Xvect-FTDNN-TrainData-I (2--3s),\\  CSML backend\end{tabular}}      & \multicolumn{1}{c|}{25.62/15.97/11.22/7.12}  & \multicolumn{1}{l|}{20.88/10.82/5.08/2.28} & \multicolumn{1}{c|}{19.45/9.96/4.80/4.15}  \\ \hline
\multicolumn{1}{|c|}{\begin{tabular}[c]{@{}c@{}}ResNet50-SE-MFB80-AM-TrainData-I (2s),\\  CS backend\end{tabular}} & \multicolumn{1}{c|}{17.67/9.89/7.12/5.75}    & \multicolumn{1}{l|}{10.22/4.12/2.17/1.56}  & \multicolumn{1}{c|}{8.86/3.73/2.18/2.07}   \\ \hline

\hline
\end{tabular}
}
\end{table*}


\subsection{Results analysis of the short utterances experiments}
Taking into account the results from Table \ref{tab:short_dur_gen_proto} obtained for modified Voxceleb1 dataset we can summarize that, while the protocol generation procedure was as close as possible to that described in \cite{XNCZ2019}, our results were somewhat different but comparable to the ones published in \cite{XNCZ2019, HE2019}. The observed differences can be attributed to the differences in test protocols. Nevertheless, synchronous testing showed significantly better quality of the proposed model for various short durations (1 sec, 2 sec and 5 sec). It should be noted that the model proposed in \cite{HE2019} also demonstrated significantly better quality on short utterances compared to the model \cite{XNCZ2019}. Unfortunately, we were unable to test the \cite{HE2019} model due to its public unavailability.

Having analyzed the results presented in 
Table \ref{tab:short_dur}, we can say that training ResNet based embedding extractors on short utterances leads to an improvement in its performance for shorter durations and degradation for full durations. This is confirmed by the results of ResNet34-MFB80-AM-TrainData-I model, trained on 1-second and 2-second segments. Thus, it is possible to slightly improve the quality of the systems for short durations in this way.

The results show that TDNN based x-vectors systems degrade more than ResNet systems on short segments test. We can observe this effect by analysing the results for Xvect-FTDNN-TrainData-I and ResNet34-MFB80-AM-TrainData-II (1s) systems on the VOiCES (dev set) and Xvect-FTDNN-TrainData-I and ResNet34-MFB80-AM-TrainData-II (2s) on VOiCES (eval set).

 The better the acoustic conditions of the dataset are the more visible relative performance degradation can be observed on it. This is confirmed by the results obtained for described x-vector and ResNet systems on VoxCeleb1-O cleaned and VOiCES eval datasets.



\section{Conclusion}
During the VOiCES challenge we noticed that the task-oriented training data preparation can significantly improve the quality of the final SR systems. But at the same time this can lead to system overfitting. 
The results of the current work allows to conclude that our x-vector based systems submitted to the VOiCES challenge are not robust for short duration test utterances and domain mismatch conditions (for example VoxCeleb1-O (cleaned) protocol).
Obtained results confirm that deep ResNet architectures are more robust and allow to improve the quality of speaker verification for both long-duration and short-duration utterances.
Our best performing system for VOiCES protocols is ResNet34 based system built on high frequency resolution MFB features. It is trained with AM-Softmax based loss function.
We should also note that utilizing the proposed U-net based VAD (instead of energy based VAD), scoring model in-domain centering and score normalization techniques provide additional performance gains for proposed SR systems.

\section{Acknowledgements}
This work was partially financially supported by the Government of the Russian Federation (Grant 08-08) and by the Foundation NTI (contract 20/18gr) ID 0000000007418QR20002.
\bibliographystyle{IEEEbib}

%

\end{document}